\documentclass[twocolumn,amsmath,amssymb,floatfix,nofootinbib,secnumarabic]{revtex4}
\pdfoutput=1

\usepackage{graphicx}
\usepackage{amsmath}
\usepackage{amssymb}
\usepackage{amsthm}
\usepackage{graphicx}
\usepackage{hyperref} 
\usepackage{multirow} 
\usepackage{wasysym} 
\usepackage[bf]{caption}
\usepackage{color}
\usepackage{epsfig}
\usepackage{feynmf}
\usepackage{verbatim} 
\usepackage{hyperref}
\begin{document}

\preprint{SU-ITP-09/36}

\setlength{\unitlength}{1mm}

\title{
\begin{flushright}
\mbox{\normalsize SU-ITP-09/36}
\end{flushright}
\vskip 15 pt
Experimental constraints on the free fall acceleration of antimatter}

\author{Daniele S. M. Alves}
\email{alves@stanford.edu}
\affiliation{Department of Physics, Stanford University, Stanford, CA 94305-4060} 

\author{Martin Jankowiak}
\email{janko@stanford.edu}
\affiliation{Department of Physics, Stanford University, Stanford, CA 94305-4060} 

\author{Prashant Saraswat}
\email{ps88@stanford.edu}
\affiliation{Department of Physics, Stanford University, Stanford, CA 94305-4060} 

\begin{abstract}
In light of recent experimental proposals to measure the free fall acceleration of 
antihydrogen in the earth's gravitational field, we investigate the bounds that 
existing experiments place on any asymmetry between the free fall of matter and 
antimatter. We conclude that existing experiments constrain any such asymmetry to be 
less than about $10^{-7}$. First we consider contributions to
the inertial masses of atoms that encode the presence of antimatter and use
precision E\"otv\"os experiments to establish the level at which they
satisfy the equivalence principle. In particular we focus on vacuum polarization
effects and the antiquark content of nucleons.  Second we consider a class of
theories that contain long range scalar and vector forces that cancel with one
another to some high precision. By construction such theories would be able to
evade detection in E\"otv\"os experiments that utilize matter while still 
allowing for a signal in antimatter experiments. Even taking such cancellation
for granted, however, we show that the radiative damping of binary pulsar 
systems constrains these forces to be significantly weaker than gravity. 
Furthermore we show that there are limits to the accuracy with which such 
cancellation can be arranged: first by determining the precision to which scalar
charges can track vector charges in the best candidate theories; and, second, by
showing that the different velocity dependence of scalar and vector forces 
necessarily introduces non-cancellation at a quantifiable level.
\end{abstract}

\maketitle

\section{Introduction} 
\label{sec:1}

  Experimentalists and theorists alike have long considered the possibility that matter and 
  antimatter fall differently in the gravitational field of the earth (e.g. see \cite{GoldmanNieto:1991,Fischler:2008zz}). Early experimental endeavors began with Fairbank, 
  who attempted to measure the differential free fall acceleration of electrons and positrons. 
  These efforts, however, did not result in any conclusive measurement because of the extreme 
  difficulty of isolating the test particles from stray electric fields.  Recently an experiment 
  has been proposed at Fermilab that aims to directly measure the free fall acceleration of 
  antihydrogen in the field of the earth, $g_{\overline{\text{H}}}$, with an expected precision of 1\% or
  better \cite{AGE}.  Likewise, another experiment (AEGIS \cite{Aegis}) to measure $g_{\overline{\text{H}}}$ 
  has been proposed at CERN.  In light of these experimental proposals, it is only reasonable to consider what sort
  of bounds existing experiments place on the inequality of $g_{\text{H}}$ and $g_{\overline{\text{H}}}$.  Although 
  some of the arguments we make can be found elsewhere in the literature, we include them here to stress the 
  point that existing experiments already place stringent bounds on any gravitational asymmetry between matter 
  and antimatter.  
  
  There are two broad classes of theoretical possibilities for how gravitational asymmetry might be realized.  The 
  first is a modification of general relativity itself.  Any such theory in which matter and antimatter gravitate
  differently will necessarily do violence to fundamental principles of general relativity and quantum field theory.  
  As such, we are not aware of any concrete, self-consistent theoretical formulation---whether well-motivated or 
  not---in which such an asymmetry exists.  Nevertheless, we can still establish bounds on any such 
  asymmetry, since existing experiments already tell us something about how antimatter gravitates.  This will
  be the subject of section \ref{sec:3}.  The essential point is that the composite nature of atoms 
  implies that precision E\"otv\"os experiments, which have been done with a variety of elements, are 
  sensitive to the gravitational  coupling of antimatter.   
    
\begin{table*}[htb]
\caption{Constraints on the E\"otv\"os parameter for various test bodies falling in the gravitational field of the earth or the sun}
\begin{center}
\begin{tabular}{| l | l | l | l | l |}
\hline
\textbf{Experiment} & \textbf{Test bodies} & \textbf{Measurement} \\ \hline
Lunar laser ranging & Earth - Moon & $\eta_{\odot,\oplus \text{-}\scriptsize{\leftmoon}} \ = (-1.0\pm1.4)\times10^{-13}$  \\ \hline
Braginsky and Panov & Al - Pt & $\eta_{\odot, \text{Al-Pt}} = (3\pm4)\times10^{-13}$  \\ \hline
E\"ot-Wash & Be - Ti & $\eta_{\oplus, \text{Be-Ti}}= (0.3\pm1.8)\times10^{-13}$ \\ \hline
E\"ot-Wash & Be - Al & $\eta_{\oplus, \text{Be-Al}}= (-1.5\pm1.5)\times10^{-13}$ \\ \hline
E\"ot-Wash & Be - Cu & $\eta_{\oplus, \text{Be-Cu}}= (-1.9\pm2.5)\times10^{-12}$ \\ \hline
\end{tabular}
\end{center}
\label{tab:1}
\end{table*} 

  The second possibility is to leave gravity itself untouched and introduce long range forces of (sub)gravitational
  strength mediated by scalar and/or vector particles.  From a theoretical perspective this ``fifth force" scenario is 
  more tractable, since in this case we have a well-defined and predictive theory that does not violate any of the
  general principles that underpin the framework of quantum field theory cum general relativity.  There are two
  distinct ways in which such forces could have evaded detection in all existing experiments.  The first is simply 
  that they could be incredibly weak, many orders of magnitude weaker than gravity.  The second
  \cite{GoldmanNieto:1991} is that such forces could be of gravitational strength but would have evaded detection 
  in all existing (matter-matter) experiments because they cancel among themselves to a sufficiently high precision.
  This latter scenario can lead to a measurable asymmetry between $g_{\text{H}}$ and $g_{\overline{\text{H}}}$, 
  since while scalar-mediated forces are universally attractive, vector-mediated forces can be either attractive 
  or repulsive, depending on the relative sign of the charges.  Thus any cancellation of these new forces in 
  matter-matter interactions will be undone when considering matter-antimatter interactions, since the vector 
  force switches from repulsive to attractive.  Thus such a theory predicts that 
  $g_{\text{H}}\ne g_{\overline{\text{H}}}$.  Bounds from existing experiments on this 
  scalar-vector scenario will be discussed in section \ref{sec:4}.  The bottom line is that composition dependence
  of free fall acceleration, which is tightly constrained by precision E\"otv\"os experiments, is generic in 
  this scenario due to the compositeness of atoms and the nature of scalar and vector interactions, both of 
  which act to spoil any would-be cancellation.  
 
\section{Experimental input}
\label{sec:2}

A number of very precise experiments have been done to measure the fractional differential acceleration, 
$\eta \equiv \Delta a/a$, of test bodies of various compositions falling in the gravitational field of the earth or 
the sun.  For E\"otv\"os experiments sensitive to the gravitational field of the sun, the most precise bounds on 
the E\"otv\"os parameter $\eta_{\odot}$ come from lunar laser ranging (LLR) experiments \cite{Williams:2004qba}, 
which measure the differential acceleration between the earth and moon towards the sun, and free torsion pendulum 
experiments performed by Braginsky and Panov using multiple aluminum and platinum test bodies \cite{Braginsky}.  
Since we are mainly interested in the free fall acceleration of antihydrogen in the \textit{earth's} gravitational field, 
the most relevant experimental input for us will be the bounds obtained by the E\"ot-Wash Group at the University of 
Washington.  Their torsion balance experiments have tightly constrained $\eta_{\oplus}$ between several pairs of 
elements \cite{Schlamminger, Su}.  It is on the basis of these bounds, which are collected in Table \ref{tab:1}, that 
we will be able to tightly constrain any asymmetry between the free fall of matter and antimatter.  

\section{Atoms have many parts}
\label{sec:3}

When considering the possibility that antimatter gravitates differently from ordinary matter, one 
is really raising the more general possibility that different forms of energy gravitate differently.  
Existing free fall experiments, which have been performed with a wide variety of elements, 
put very stringent limits on any such non-universality of gravity, since the fractional contributions
of various forms of energy to the inertial masses of atoms---nuclear binding energies, atomic binding
energies, kinetic energies of the constituents, etc.---vary from element to element.  What can these
experiments tell us about how antimatter gravitates?  The essential point is that nuclei and atoms are 
composite states.  Although one can make a distinction between matter and antimatter at the level of 
quarks and electrons, that distinction is blurred when one considers bound states like nuclei and atoms. And because antimatter plays a quantifiable role in the physics of nuclei and atoms by contributing to 
their inertial masses, precision E\"otv\"os experiments utilizing matter continue to be relevant when 
considering the possibility of gravitational asymmetry between matter and antimatter. 

In particular we will focus on two ways in which antimatter enters the physics of nuclei and atoms.  First, 
in sections \ref{31} and \ref{32} we will consider contributions to the inertial masses of nuclei and atoms due to 
vacuum polarization effects.  Since these effects reflect the screening of electric charges by virtual pairs of
electrons and positrons, we interpret these contributions to the inertial masses of nuclei and atoms
as encoding their antimatter content.  Second, in section \ref{33} we will consider the sea antiquark 
content of nucleons as established by deep inelastic scattering experiments. 
In both cases we will quantify the degree to which existing E\"otv\"os experiments require these forms of 
energy to satisfy the equivalence principle.  We then make the assumption that 
any deviation of $g_{\text{H}}$ from $g_{\overline{\text{H}}}$ would manifest
itself as a violation of the equivalence principle in these forms of energy at
the same level.  This reasoning will then allow us to place bounds on
$|g_{\text{H}}\!-\!g_{\overline{\text{H}}}|/g_{\text{H}}$.  
It remains an interesting challenge to see whether it is possible to construct a theory for which the resulting bounds would not hold.
Such a theory would require the effective gravitational coupling of antimatter as probed by fermion loops and sea antiquarks
to be decoupled from the gravitational coupling of antihydrogen.  In the absence of such a theory, however,
our task is to establish the consequences of our basic assumption.  

Having outlined our approach, it remains to quantify the effects we are interested in.
In section \ref{31} we consider the Lamb shift in atoms and its implications for the universality of gravity.  
In section \ref{32} we consider the analogous and much larger effect in the electrostatic self-energies of nuclei.  
Finally in section \ref{33} we quantify the antiquark content of nucleons as well as the antimatter fractions of atoms,
which will allow us to place further constraints on any gravitational asymmetry between matter and antimatter.

\subsection{Lamb shift}
\label{31}

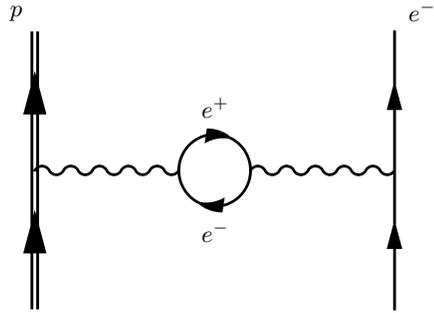
\begin{figure}[tb]
\centering
\begin{fmffile}{LambShiftx}  
\begin{fmfgraph*}(60,37)   
\fmfleft{p1,p2}  
\fmfright{e1,e2}
\fmf{dbl_plain_arrow,straight}{p1,vp,p2} 
\fmf{fermion,straight}{e1,ve,e2}  
\fmffreeze   
\fmf{photon}{vp,v1}
\fmf{photon}{v2,ve}
\fmf{fermion,label=$e^+$,left}{v1,v2}
\fmf{fermion,label=$e^-$,left}{v2,v1}
\fmflabel{$p$}{p2}   
\fmflabel{$e^-$}{e2}     
\end{fmfgraph*} 
\end{fmffile}\caption{Vacuum polarization contribution to the Lamb shift}
\label{fig:1}
\end{figure}

Among the most precisely verified predictions of quantum electrodynamics is the Lamb shift in hydrogenlike atoms.  
Historically the term ``Lamb shift" refers to the splitting between the 2$s_{1/2}$ and 2$p_{1/2}$ energy levels in the 
hydrogen atom; here we use it to refer to any correction to hydrogenlike energy levels from the values obtained by 
solving the Dirac equation for the Coulomb potential.  One contribution to the Lamb shift is given by the vacuum 
polarization diagram of Figure \ref{fig:1}.  The electron loop in this diagram contributes to the running of the QED coupling 
constant at energies above the electron mass $m_e$, which results in an effective electrostatic potential, the Uehling 
potential, that differs from the usual Coulomb $1/r$ potential at distances shorter than $m_e^{-1}$.  This modification 
of the Coulomb potential at short distances can be interpreted as screening of the nuclear charge by pairs of virtual 
electrons and positrons.  In an abuse of terminology, we will refer to the energy shift due to this effect as the Lamb shift, 
even though it constitutes only a fraction of the total Lamb shift (about 2--30\% of the total depending on $Z$). 
The Lamb shift for the $n^{\textup{th}}$ energy level is given by  
 \begin{equation}
 E_{\text{Lamb}}^{l=0} = -\frac{\alpha (Z \alpha)^4}{\pi n^3} F(Z\alpha) \; m_e
 \end{equation}
where $Z$ is the atomic number and $F(Z\alpha)$ varies slowly with $Z$ (from about .25 to 1 as $Z$ goes from 1 to 
100) \cite{johnson85}.  The total Lamb shift has been measured for high-$Z$ hydrogenlike atoms, up to uranium 
($Z\!=\!92$) \cite{gumberidze05}.  These measurements confirm the predictions of QED in the strong-field regime.

The strong $Z$-dependence of the Lamb shift implies that the fractional contribution of this form of energy to 
the inertial masses of atoms varies appreciably from element to element.  Thus the accuracy to which the 
Lamb shift satisfies the equivalence principle can be constrained by precision E\"otv\"os experiments that 
utilize a variety of elements.  For example, for beryllium the Lamb shift is a fraction $\sim 4 \times 10^{-14}$ 
of the total mass, while for titanium the fraction is $\sim 9 \times 10^{-12}$.  Of course these atoms are much 
more complicated than hydrogen, but for our purposes it is a good approximation to calculate the energy 
shift of the 1s electrons as if the atom were hydrogenlike.  Using the experimental input from section \ref{sec:2}, 
this reasoning yields a bound at the percent level:
\begin{equation}
|\eta_{\oplus, \text{Be-Ti}}|=\Delta\left(\frac{E_{\text{Lamb}}}{m_{\text{atom}}} \right)\frac{|g-g_{{\text{Lamb}}}|}{g}
 \lesssim 10^{-13}
\end{equation} 
\vspace{-5mm}
\begin{equation}
 \Rightarrow \frac{|g-g_{{\text{Lamb}}}|}{g} \lesssim 10^{-2}
\end{equation}
Therefore the Lamb shift contribution to the inertial mass satisfies the equivalence principle to about one part in $10^2$. 
We interpret the diagram in Figure \ref{fig:1} as (perturbatively) encoding physical effects of antimatter in the atom.
For that reason we expect $|g\!-\!g_{{\text{Lamb}}}|/g$ to be related to $|g_{\text{H}}\!-\!g_{\overline{\text{H}}}|/g_{\text{H}}$ by an 
$\mathcal{O}(1)$ factor.  This reasoning then yields a bound $|g_{\text{H}}\!-\!g_{\overline{\text{H}}}|/g_{\text{H}}\lesssim 10^{-2}$. 
As we shall see in the next section, there is an analogous and much larger effect in the electrostatic self-energies of nuclei.  

\subsection{Electrostatic self-energy of the nucleus}
\label{32}

\begin{figure}[tb]
\centering
\begin{fmffile}{NuclearLamb}  
\begin{fmfgraph*}(80,35)   
\fmfbottom{q1,q2}  
\fmf{dbl_plain}{q1,v1,v2,q2}   
\fmfforce{(.43w,.643h)}{v3}
\fmfforce{(.57w,.6435h)}{v4}
\fmf{fermion,right,label=$e^-$}{v3,v4}
\fmf{fermion,right,label=$e^+$}{v4,v3}
\fmffreeze   
\fmf{photon,left=.5}{v1,v3}
\fmf{photon,left=.5}{v4,v2}    
\end{fmfgraph*} 
\end{fmffile}
\caption{Loop contribution to the electrostatic self-energy of the nucleus}
\label{fig:2}
\end{figure}
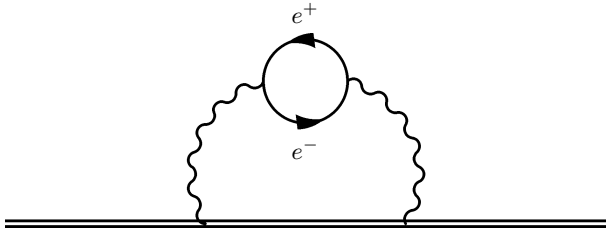

Whereas the Lamb shift typically constitutes an $\mathcal{O}(10^{-14}\!-\!10^{-12})$ fraction of an atom's inertial mass, 
electron loops make a much larger contribution to the electrostatic self-energy of the nucleus \cite{Polchinski:2006gy}. 
The classical electrostatic self-energy of the nucleus scales like
\begin{align}
E_{\text{EM}} \simeq - \frac{3}{5} \frac{\alpha Z (Z-1)}{A^{1/3} R_0} \approx \frac{0.72 Z (Z-1)} {A^{1/3}} \, (\text{MeV})
\end{align}
where $R_0=1.2\hspace{0.03 in} \text{fm}$.  The leading correction (Figure \ref{fig:2}) to this electrostatic energy comes from 
the insertion of a vacuum polarization loop analogous to that in the Lamb shift.  This contribution amounts to a relative 
correction to the electrostatic self-energy
\begin{align}
\frac{E_{\text{Loop}}}{E_{\text{EM}}} \simeq \frac{\alpha}{4 \pi} \log(m_{\text{e}}^2 R_{\text{nuc}}^2) \approx 10^{-3}
\end{align}

\!\!There are of course additional (and potentially much larger) corrections coming from QCD and quark loops. 
Nonetheless, the above contribution from QED alone, which constitutes an $\mathcal{O}(10^{-6})$ fraction of 
the inertial mass of the nucleus, is enough to provide a significant bound:
\begin{equation}
|\eta_{\oplus, \text{Be-Ti}}|=\Delta\left(\frac{E_{\text{Loop}}}{m_{\text{atom}}} \right)\frac{|g-g_{\text{Loop}}|}{g} \lesssim 10^{-13}
\end{equation}
\vspace{-5mm}
\begin{equation}
\Rightarrow  \frac{|g-g_{\text{Loop}}|}{g} \lesssim 10^{-7}
\end{equation}
since the difference in the fractional contribution of $E_{\text{Loop}}$ to the inertial masses of beryllium and titanium is 
$\Delta\!\left(\frac{E_{\text{Loop}}}{m_{\text{atom}}} \right) \approx 10^{-6}$.  Therefore the electron loop contribution 
to the electrostatic self-energy of the nucleus satisfies the equivalence principle to about one part in $10^7$.  In exact analogy
to the previous section, we expect $|g\!-\!g_{\text{Loop}}|/g$ to be related to $|g_{\text{H}}\!-\!g_{\overline{\text{H}}}|/g_{\text{H}}$ 
by an $\mathcal{O}(1)$ factor.  This reasoning then yields a bound $|g_{\text{H}}\!-\!g_{\overline{\text{H}}}|/g_{\text{H}}\lesssim 10^{-7}$.

\subsection{Antiquarks in nucleons}
\label{33}

Deep inelastic scattering experiments have thoroughly established that the constituents of the proton and neutron 
include the antiquarks $\overline{u}, \overline{d}$, and $\overline{s}$.  Furthermore, the corresponding parton distribution
functions have been measured at the percent level.  We can take the moment

\begin{equation}
\int_0^1x\{\overline{u}(x)+\overline{d}(x)+\overline{s}(x)\} dx \approx 0.1
\end{equation}
as characterizing the antiquark energy fraction of a nucleon.  Making use of the fact that different elements carry different 
nuclear (and therefore antimatter) energy fractions simply because the ratio of nucleons to electrons, as well as the 
nuclear binding energy, varies from element to element, we can establish a bound on $|g-g_{\overline{\text{q}}}|/g$.  Consider 
two different elements A and B with inertial masses $m_{\text{A}}$ and $m_{\text{B}}$, respectively. The inequality of 
$g$ and $g_{\overline{\text{q}}}$ will drive the E\"otv\"os parameter  $\eta_{\oplus\text{,A-B}}$ away from zero:

\begin{equation}
|\eta_{\oplus\text{,A-B}}|=F_{\overline{\text{q}}}|F_{\text{N}}^{\text{A}}-F_{\text{N}}^{\text{B}}|\frac{|g-g_{\overline{\text{q}}}|}{g}
\end{equation}
where $F_{\overline{\text{q}}} \sim 0.1$ is the antiquark energy fraction carried by a nucleon, and $F_{\text{N}}^{\text{A}}$ and 
$F_{\text{N}}^{\text{B}}$ are the nuclear mass fractions of elements $A$ and $B$, respectively.  Since we have 
$|F_{\text{N}}^{\text{Be}}-F_{\text{N}}^{\text{Ti}}| \sim 10^{-3}$, we obtain a bound $|g-g_{\overline{\text{q}}}|/g \lesssim 10^{-9}$.  Therefore 
the antiquarks in atoms satisfy the equivalence principle to about one part in $10^9$.  Since we are assuming that
$|g-g_{\overline{\text{q}}}|/g$ is related to $|g_{\text{H}}\!-\!g_{\overline{\text{H}}}|/g_{\text{H}}$ by an $\mathcal{O}(1)$ factor,
this yields a bound $|g_{\text{H}}\!-\!g_{\overline{\text{H}}}|/g_{\text{H}}\lesssim 10^{-9}$.

\section{A fifth force cancelled by a sixth force}
\label{sec:4}

Here we consider the possibility that there exist long range forces of (sub)gravitational strength mediated
by scalar and/or vector particles.  Since any new long range scalar or vector force by itself is constrained to 
be extremely weak (see e.g. \cite{Adelberger:2003zx, Dolgov:1999gk}), we consider situations in which a
scalar-mediated force is approximately canceled by a vector-mediated force \cite{GoldmanNieto:1991}.  
As discussed in the introduction, this approximate cancellation would still allow for a large deviation from 
$g_{\text{H}} = g_{\overline{\text{H}}}$, since the force due to vector exchange becomes attractive when the test 
particle is an antiparticle.  

In the following we leave aside any questions about the theoretical plausibility of this scenario---e.g. the level of 
fine-tuning required or whether some symmetry might enforce approximate cancellation---and establish bounds 
on the theory as given.  In section \ref{41} we obtain bounds on $|g_{\text{H}}\!-\!g_{\overline{\text{H}}}|/g_{\text{H}}$ 
by considering the radiative damping of binary pulsar systems.  
These bounds have the virtue of holding irrespective of any precise cancellation. In section \ref{42} we investigate 
scalar-vector scenarios in which approximate cancellation is possible and quantify the degree to which that 
cancellation fails.  In section \ref{43} we demonstrate that scalar forces cannot be arranged to exactly cancel 
against vector forces as a consequence of their different velocity dependence and 
quantify the degree of non-cancellation which necessarily results.  In both sections \ref{42} and \ref{43} we use the computed degree of non-cancellation in conjunction with input from precision E\"otv\"os 
experiments to place bounds on $|g_{\text{H}}\!-\!g_{\overline{\text{H}}}|/g_{\text{H}}$.  

Note that although the physical effects considered in section \ref{sec:3} have bearing on the composition 
dependence of scalar- and vector-mediated forces (since they contribute to the renormalization of
the effective coupling constants), we do not explicitly explore this connection here, since the arguments 
presented below are better suited to constraining general combinations of long range scalar and vector forces.

\subsection{Radiative damping of binary pulsar systems}
\label{41}

Even supposing that scalar and vector forces could somehow
be fine-tuned so as to evade detection in precision E\"otv\"os experiments, no 
amount of fine-tuning can circumvent the fact that particles charged under long range forces can radiate
energy.  In particular binary pulsar systems can radiate off enough 
energy in the form of scalar or vector waves to modify their orbital decay at an observable level, 
provided that the range of the scalar or vector force is somewhat larger
than their orbital period, $\lambda \gtrsim P_b$, where typically $ P_b \approx 10^{12} \textup{m} \approx 10^{5} R_{\oplus}$.  
The possibility of using binary pulsar systems to constrain
long range scalar and vector forces was considered in \cite{Krause:1994ar},
since which time the number of precisely measured binary systems has increased considerably.  
We consider two separate cases: (i) the vector couples to baryon number $B$; and (ii) the vector
couples to lepton number $L$.  In both cases the scalar force is assumed to cancel
against the vector force, so that both forces couple to the same charge with identical strengths. By
appealing to recent observational input we will be able to tightly constrain these two scenarios.

Strong bounds on the baryon number case can be obtained by considering dipole radiation. 
The ratio of the energy loss due to dipole radiation from a vector interaction, $\langle \dot{E}_\text{V}\rangle$,
to the energy loss due to gravitational quadrupole radiation as predicted by general relativity, 
$\langle \dot{E}_\text{GR}\rangle$, is given by
\begin{equation}
\label{dipole/quadrupole vector}
\frac{\langle \dot{E}_\text{V}\rangle}{\langle \dot{E}_\text{GR}\rangle} = 
\chi(m_\text{V}, \epsilon) \frac{\alpha_\text{V}}{\alpha_\text{GR}} \left[ \Delta\left(\frac{B}{\mu}\right)\right]^2\frac{1}{a^2\omega^2}
\end{equation}
where $\alpha_\text{V}\equiv g_\text{V}^2/4\pi$ characterizes the strength of the vector interaction, 
$\alpha_\text{GR}\equiv Gm_{\text{H}}^2$ with $m_{\text{H}}$ the mass of the hydrogen atom, 
$B$ is the star's baryon number, $\mu$ is the star's mass in units of $m_{\text{H}}$, 
$a$ is the semimajor axis of the relative orbit, $\omega=2\pi/P_b$ is the characteristic frequency of the system, 
and $\chi(m_V, \epsilon)$ is a geometric factor that depends on the orbital eccentricity $\epsilon$ and the 
mass of the vector particle $m_\text{V}$ \cite{Krause:1994ar}.  Here the term $\Delta(B/\mu)$ characterizes the size of the baryonic
dipole moment, and the factor of $a^2\omega^2$ reflects the fact that this is a ratio of dipole radiated power 
to quadrupole radiated power. An analogous expression holds for scalar radiation
\begin{equation}
\label{dipole/quadrupole scalar}
\frac{\langle \dot{E}_\text{S}\rangle}{\langle \dot{E}_\text{GR}\rangle} =
\frac{1}{2} \chi'(m_\text{S}, \epsilon) \frac{\alpha_\text{S}}{\alpha_\text{GR}} \left[ \Delta\left(\frac{B}{\mu}\right)\right]^2\frac{1}{a^2\omega^2}
\end{equation}
where the geometric factor $\chi'(m_\text{S}, \epsilon)$ reduces to $\chi(m_\text{V}, \epsilon)$ in the limit $m_\text{V,S} \rightarrow 0$.

\begin{table}[b]
\caption{Observed and inferred orbital parameters of J1141-6545 \cite{Bhat:2008ck}}
\begin{center}
\begin{tabular}{| l | l | l | l | l |}
\hline
\textbf{Parameter} & \textbf{Measured value} \\ \hline
Orbital Period $P_b$ & $0.1976509593(1)$ days \\ \hline
Eccentricity $\epsilon$ & $.171884(2)$ \\ \hline
Advance of Periastron $\dot{\omega}_{\text{GR}}$ & $5.3096(4)$ deg yr$^{\textrm{-1}}$ \\ \hline
Observed Period Derivative $\dot{P}_{b}^{\text{obs}}$ & $-0.403(25) \times 10^{-12}$  \\ \hline
Intrinsic Period Derivative $\dot{P}_{b}^{\text{intrinsic}}$ & $-.401(25) \times 10^{-12}$ \\ \hline
Ratio of $\dot{P}_{b}^{\text{intrinsic}}$ to GR prediction & $1.04(6)$  \\ \hline
\end{tabular}
\end{center}
\label{tab:2}
\end{table}

In order to constrain the energy loss due to scalar and vector dipole radiation we relate the observed change
of the orbital period (after correcting for various kinematics effects \cite{Bhat:2008ck}) to the GR prediction:
\begin{equation}
 \label{delta}
 \frac{\langle \dot{E}_\text{S}\rangle+\langle \dot{E}_\text{V}\rangle}{\langle \dot{E}_\text{GR}\rangle}=
 1-\frac{\dot{P}^{\text{GR}}_b}{\dot{P}^{\text{intrinsic}}_b}
 \end{equation}
which is valid for $\dot{E}_\text{V,S} \ll \dot{E}_\text{GR}$.  Going to the massless limit $m_\text{V,S} \rightarrow 0$ 
(which is a good approximation whenever the range of the scalar and vector interactions is somewhat larger 
than the orbital period), letting $\alpha_\text{S} = \alpha_\text{V} \equiv \alpha_\text{SV}$, and rewriting equations (\ref{dipole/quadrupole vector})-(\ref{delta})
in terms of the orbital period $P_b$ and the advance of periastron $\dot\omega_{\text{GR}}$ \cite{Krause:1994ar}, 
we obtain a bound
\vspace{-3mm}
\begin{equation}
\label{pulsar bound}
\frac{\alpha_\text{SV}}{\alpha_\text{GR}}\leq\frac{16}{15\pi}f(\epsilon)\dot\omega_{\text{GR}}P_b\left(1-\frac{\dot{P}^{\text{GR}}_b}{\dot{P}^{\text{intrinsic}}_b}\right) \left[ \Delta\left(\frac{B}{\mu}\right)\right]^{-2}
\end{equation}
where $f(\epsilon)$ is a monotonic function that ranges from $1$ to $2.95$ as $\epsilon$ varies from $0$ to $1$.
As is clear from (\ref{pulsar bound}), the best bounds will
come from binary systems that have large baryonic dipole moments.  For that reason the binary system J1141-6545,
which consists of a neutron star with a white dwarf companion, is a good candidate for our purposes.  It has a large baryonic dipole
moment because neutron stars have significant gravitational binding energy ($B/\mu\!\approx \!1.1$) while white dwarves
have negligible gravitational binding energy ($B/ \mu \! \approx \!1$). Using (\ref{pulsar bound}) and the observational input 
of \cite{Bhat:2008ck}, summarized in Table \ref{tab:2}, we arrive at a bound $|g_{\text{H}}\!-\!g_{\bar{\text{H}}}|/g_{\text{H}}\lesssim10^{-4}$.

Bounds for the lepton number case can be obtained in the same way, with $B$ replaced by $L$ in (\ref{pulsar bound}).
Here a system consisting of two neutron stars does not provide a good bound, since neutron stars carry few electrons.  
That neutron stars are electron poor, however, becomes useful when considering a system consisting of a neutron star and 
white dwarf, since for the white dwarf $L/\mu \approx 0.5$, which results in a large leptonic dipole moment.  Appealing
to the observational input from J1141-6545 we arrive at a bound $|g_{\text{H}}\!-\!g_{\bar{\text{H}}}|/g_{\text{H}}\lesssim10^{-5}$.  

Bounds for the baryon number case can also be obtained by considering quadrupole radiation,
since even in the case when the binary system has $\Delta(B/\mu) = 0$, there is a nonvanishing 
quadrupole moment that changes in time.   
When the scalar or vector forces are of gravitational strength, the energy loss due to scalar or vector quadrupole
radiation is comparable to that due to gravitational wave emission, since they are all sourced by essentially the
same multipole moment.  Since the GR prediction accounts for the observed orbital decay in the binary pulsar 
system B1913+16 to within about .3\% \cite{Weisberg:2004hi}, we obtain a bound
$|g_{\text{H}}\!-\!g_{\overline{\text{H}}}|/g_{\text{H}}\lesssim10^{-3}$.  

Thus radiative damping of binary pulsar systems alone places tight bounds on 
$|g_{\text{H}}\!-\!g_{\bar{\text{H}}}|/g_{\text{H}}$ in the scalar-vector
scenario.  These bounds are robust and can only be evaded (while simultaneously keeping experiments sourced by the earth 
relevant) by requiring the range of the scalar and vector forces to sit somewhere in the window
$R_{\oplus} \lesssim \lambda \lesssim 10^{5} R_{\oplus}$, which corresponds to a mass range
$10^{-19} \text{ eV} \lesssim m \lesssim 10^{-13} \text{ eV}$.  Quite interestingly for scalar particles this range of masses
can be probed by a super-radiant instability of rotating black holes \cite{Arvanitaki:2009fg}.

\subsection{Scalars charges are not vector charges}
\label{42}

Let us return to the possibility of approximate cancellation between scalar and vector forces in matter-matter interactions.   
For concreteness, let us first consider a scalar particle coupled to the 
trace of the energy momentum tensor $T_{\ \mu}^{\mu}$ (so that the scalar couples to mass) and a vector particle 
coupled to baryon number $B$, where the strengths of both couplings have been 
adjusted so as to achieve approximate cancellation and both forces have the same range $\lambda \!\gtrsim \!R_{\oplus}$.  
Approximate cancellation is possible, since across the periodic table
the ratio $B/\mu$ (where $\mu$ is the inertial mass in atomic mass units) is approximately constant: $B/\mu \approx 1$.  
The typical variation from element to element is $\Delta(B/\mu) \approx \mathcal{O}(10^{-3}-10^{-4})$.  This cancellation 
cannot be made exact, however, since there is no mechanism by which the scalar can couple precisely to $B$.  Thus 
$\eta$ between any two given elements will be nonzero:
\begin{equation}
|\eta| = \Delta\left(\frac{B}{\mu}\right)\frac{|g_{\text{H}}-g_{\overline{\text{H}}}|}{2g_{\text{H}}}.
\end{equation}
Using the experimental input from section \ref{sec:2}, this reasoning yields a bound 
$|g_{\text{H}}-g_{\overline{\text{H}}}|/g_{\text{H}} \lesssim 10^{-8}$. If the vector instead couples to lepton number $L$, one 
obtains a more stringent bound $|g_{\text{H}}-g_{\overline{\text{H}}}|/g_{\text{H}} \lesssim 10^{-10}$, since the ratio $L/\mu$ 
varies more strongly across the periodic table: $\Delta(L/\mu) \approx \mathcal{O}(10^{-1}\!-\!10^{-2})$.  Bounds of similar 
order of magnitude or better will hold for other scenarios \cite{Adelberger}, e.g. if the scalar couples primarily to glue 
($\mathcal{L} \! \supset \! \phi \thinspace \textup{Tr} \thinspace F_{QCD}^2$).  Even allowing for the possibility that one 
introduces additional adjustable parameters corresponding to additional interactions (e.g. non-renormalizable vector 
interactions or a scalar coupling to the photon field strength squared) to improve the cancellation between the elements 
that have been tested in E\"otv\"os experiments, it is unlikely that the degree of cancellation will be such that the above 
bounds on $|g_{\text{H}}-g_{\overline{\text{H}}}|/g_{\text{H}}$ will be substantially weakened---even allowing for incredible 
fine-tuning.  In any case we need not consider the entire spectrum of possible interactions, since in the next section we will 
give a robust argument that does not depend on the particular form of the interactions.  

\subsection{Bounds from the velocity dependence of scalar and vector forces}
\label{43}

Further limits on any cancellation between scalar and vector interactions can be obtained by considering 
how the corresponding forces transform differently under Lorentz boosts.  In particular for two pointlike particles the magnitude 
of the inverse-square force mediated by the vector is larger than that mediated by the scalar by a relative factor $u_1 \cdot u_2$, 
where the $u_i$ are the four-velocities of the particles \cite{Macrae}.  Hence any cancellation between the two forces in the static 
limit will be undone in the nonstatic case.  Therefore in the case where the vector couples to some linear combination of $B$ and 
$L$ (and the scalar interaction somehow tracks the same linear combination to some high precision), the motion of nucleons 
within the nucleus and electrons within the atom places limits on the precision of the would-be cancellation, since the average 
velocities of the nucleons and electrons will vary from element to element.  Within the Fermi gas model we can calculate the average
kinetic energy of a nucleon, inside a nucleus consisting of $Z$ protons and $N$ neutrons, to be  
\begin{equation}
\langle E_{\textup{kin}}\rangle= (31 \textup{\ MeV}) \frac{Z^{5/3}+N^{5/3}}{(Z+N)^{5/3}}
\end{equation}
\noindent This corresponds to an average Lorentz boost factor 
\begin{equation}
\langle \gamma-1\rangle = \langle\frac{1}{2}v^2\rangle = (3 \times 10^{-2}) \frac{Z^{5/3}+N^{5/3}}{(Z+N)^{5/3}}
\end{equation}
\noindent After averaging over the velocities of the nucleons within a nucleus the effect of the factor $u_1\! \cdot u_2$ is to 
introduce a relative factor of $\langle \gamma \rangle$ between the magnitudes of the scalar- and vector-mediated forces.  
Note that the quantity $\frac{Z^{5/3}+N^{5/3}}{(Z+N)^{5/3}}$ varies at the $10^{-3}$ level between elements.  Combined with
the experimental input from section \ref{sec:3}, this yields a bound $|g_{\text{H}}\!-\!g_{\overline{\text{H}}}|/g_{\text{H}} \!\lesssim \!10^{-7}$ 
for the case where the vector couples primarily to $B$.  The same reasoning applies in the case where the vector couples primarily
to $L$.  Here one obtains a bound $|g_{\text{H}}\!-\!g_{\overline{\text{H}}}|/g_{\text{H}}\! \lesssim \!10^{-9}$, which is more stringent 
than in the former case because electron velocities vary more from element to element than nucleon velocities.  These bounds are 
robust and cannot be simply evaded by postulating further scalar and/or vector interactions and further fine-tuning.  Since these 
bounds are derived by considering the kinetic energies of nucleons and electrons within atoms, one might imagine that they could 
in principle be weakened somewhat with the addition of scalar couplings to field strengths squared, since these approximately track 
the binding energy and therefore (by virial theorems) the kinetic energies of the constituent nucleons and electrons \cite{GoldmanNieto:1991}. 
But since the ratio of nuclear (and electronic) binding energy to kinetic energy varies significantly from element to element, 
the inclusion of such interaction terms does nothing to weaken the above bounds.

\begin{table*}[htb]
\caption{Summary of bounds.  See section \ref{sec:3} for details on bounds on the scenario where GR is modified and section \ref{sec:4} for details 
on bounds on the scalar-vector scenario.}
\begin{center}
\begin{tabular}{| l | l | l | l | l |}
\hline
\textbf{Scenario} & \textbf{Argument} & \textbf{Bound on }$|g_{\text{H}}\!-\!g_{\overline{\text{H}}}|/g_{\text{H}}$ \\ \hline
\multirow{2}{*}{Modification of GR}
& Lamb shift & $\lesssim  10^{-2}$ \\ 
& Electrostatic self-energies of nuclei & $\lesssim 10^{-7}$  \\ 
& Antiquarks in nucleons & $\lesssim 10^{-9}$  \\ \hline
\multirow{2}{*}{Scalar-vector}
& Radiative damping of binary systems & $\lesssim 10^{-4}$  \\
& Scalar charges are not vector charges & $\lesssim 10^{-8}$  \\
& Velocity dependence & $\lesssim 10^{-7}$  \\ \hline
\end{tabular}
\end{center}
\label{tab:3}
\end{table*}

The above considerations are just one of the many effects that contribute to the renormalization of the effective
scalar and vector charges of atoms.  The implication of these effects is that even if cancellation could be achieved at the level of 
electrons, protons, and neutrons, that cancellation would necessarily 
be undone as one descends to the (relevant) effective theory in which atoms are the degrees of freedom.  As a consequence 
composition dependence is generic in the scalar-vector scenario.

\section{Summary}
\label{sec:5}

The discussion in this paper was motivated by recent experimental proposals to test for violations of the equivalence principle 
in the free fall acceleration of antihydrogen in the gravitational field of the earth.  Focusing our attention
on two different scenarios for how such gravitational asymmetry between matter and antimatter might
be realized, we established a number of strict bounds, which are collected in Table \ref{tab:3}.
\begin{acknowledgments}
The authors acknowledge the support of the Stanford Institute for Theoretical Physics and
the hospitality of the theory group at the University of Oxford, where part of this work was completed.  
We are also grateful to Nathaniel Craig, Savas Dimopoulos, Sergei Dubovsky, Mark Kasevich and Michael Peskin for useful discussions.
This work was supported by the NSF under grant PHY-0244728.
\end{acknowledgments}

\bibliographystyle{unsrt}

\end{document}